\documentclass[prl,nobalancelastpage,twocolumn,superscriptaddress,nolongbibliography]{revtex4-2}
\pdfoutput=1

\usepackage{color,amsthm,amsmath,amsxtra,amsfonts,dsfont,graphicx,bm,amssymb}
\usepackage[colorlinks=true,linkcolor=blue, citecolor=blue, urlcolor=blue, bookmarks]{hyperref}
\usepackage{centernot}
\usepackage[dvipsnames]{xcolor}
\usepackage{graphicx}
\usepackage{tikz}
\usepackage{braket}

\def\Id{{\openone}}
\newcommand{\be}{\begin{equation}}
\newcommand{\ee}{\end{equation}}
\newcommand{\bea}{\begin{eqnarray}}
\newcommand{\eea}{\end{eqnarray}}
\newcommand{\bse}{\begin{subequations}}
\newcommand{\ese}{\end{subequations}}

\theoremstyle{plain}

\newtheorem{thm}{Theorem}[section]
\newtheorem{prop}{Proposition}[section]

\newtheorem{defn}{Definition}[section]

\newtheorem{example}{Example}[section]

\newcommand{\prlsection}[1]{{\em {#1}.---~}}

\begin{document}

\title{Quantum Circuits assisted by LOCC: Transformations and Phases of Matter}

\begin{abstract}
We introduce deterministic state-transformation protocols between many-body quantum states which can be implemented by low-depth Quantum Circuits (QC) followed by Local Operations and Classical Communication (LOCC). We show that this gives rise to a classification of phases in which topologically-ordered states or other paradigmatic entangled states become trivial. We also investigate how the set of unitary operations is enhanced by LOCC in this scenario, allowing one to perform certain large-depth QC in terms of low-depth ones.

\end{abstract}

\author{Lorenzo \surname{Piroli}}
\author{Georgios \surname{Styliaris}}
\author{J.~Ignacio \surname{Cirac}}
\affiliation{Max-Planck-Institut f{\"{u}}r Quantenoptik, Hans-Kopfermann-Str. 1, 85748 Garching, Germany}
\affiliation{Munich Center for Quantum Science and Technology (MCQST), Schellingstr. 4, D-80799 München, Germany}

\maketitle

% --------------------------------------------------------------
% --------------------------------------------------------------

Recently, we have witnessed the formation of close connections between Quantum Information Theory (QIT) and Quantum Many-Body Physics (QMBP). Among already established ones, a potential area of common interest is the classification of quantum states and operations. For instance, in QIT one is interested in states that are related by local operations and classical communication (LOCC), since entanglement is seen as a resource and those operations do not increase it~\cite{horodecki2009quantum}. In QMBP, instead, one is interested in the phases of matter which are dictated by local (unitary) transformations~\cite{chen2010local,hastings2013classifying,zeng2015quantum,zeng2015gapped,chiu2016classification}, since those are the ones typically occurring in nature. Despite the apparent similarities, the goals and methods in these fields are very different. First, the notion of locality is not the same. In QIT, there is no underlying geometry, so it usually refers to operations that act on a qubit (or subset of qubits), independent of their location. In QMBP, instead, there is an underlying geometry (typically a lattice), and locality refers to operations or Hamiltonians acting on subsystems close to each other. In addition, in QIT measurements and communication are allowed, while these are not traditionally considered in QMBP scenarios (although, recently, a lot of attention has been devoted to unitary dynamics in many-body systems subject to repeated  measurements, see e.g.~\cite{yaodong2018quantum,skinner2019measurement,fan2020self,jian2020measurement, choi2020quantum,gullans2020dynamical,ippoliti2021entanglement}).

The advent of Noisy Intermediate-Scale Quantum (NISQ) devices~\cite{preskill2018quantum} has attracted the interest of both communities, providing a unique framework to share methodologies and pursue common goals. Those devices operate quantum circuits (QC), where quantum gates act on nearest neighbors according to some lattice geometry. Additionally, single qubit measurements can be performed, and local gates applied depending on the outcomes. Thus, it is very natural to consider the classification of states, phases of matter, or actions in general under a new paradigm that includes both the local operations appearing in QMBP and the LOCC of QIT. This has clearly practical motivations: for instance, accounting for LOCC could improve the efficiency of recent preparation protocols of topologically ordered states with quantum devices~\cite{satzinger2021realizing}, possibly scaling to larger number of qubits. At the fundamental level, this problem provides a  common ground for QIT and QMBP, with the potential to  motivate a fruitful cross-fertilization of ideas.

In this work we establish a framework to address this question. We consider state transformations and unitary operations via finite-depth QC assisted by LOCC, highlighting how this leads to new possibilities. We show that paradigmatic examples, such as the Toric-Code (TC)~\cite{kitaev2003fault}, the GHZ and W states~\cite{greenberger1989bell,dur2000three}, appear in the trivial phase. Furthermore, we provide a full classification of phases in 1D in the context of Matrix Product States (MPS)~\cite{fannes1992finitely,perez2007matrix,cirac2017matrix_op}, extending that analyzed in \cite{chen2010local,schuch2011classifying}. For operations, LOCC enhance the potential of QC, enabling the implementation of unitary transformations that would require complex QC, which may become useful in the design of future quantum computers.

\prlsection{Quantum circuits and LOCC}
We consider spins arranged over a $N\times \cdots N=:\Lambda_{N,D}$ regular lattice in $D$ spatial dimensions. The associated Hilbert space is $H=H_d^{\otimes M}$, with $M=N^D$ spins. The local space is $H_d$, has dimension $d$, and we will call $\{|0\rangle,\ldots,|d-1\rangle\}$ the computational basis. We denote by ${\cal U}$ the set of unitary transformations acting on the spins. We begin by introducing the class of QC, as the operators $V\in {\cal U}$ that are decomposed as a sequence of unitaries $ V =  V_{\ell} \ldots V_2 V_1$, where each ``layer'' $V_{n}$ contains quantum gates acting on disjoint pairs of nearest-neighbor spins~\footnote{Often, gates are restricted to belong to some finite (universal) gate set. Here, since we focus on properties related to locality, we instead consider as a valid gate any two-qudit unitary operator acting on nearest neighbors.}. We call $\ell$ the circuit depth.
\begin{defn}
[Depth-$\ell$ quantum circuits] ${\cal QC}_{\ell}\subset {\cal U}$ is the set of unitaries that can be  expressed as quantum circuits of depth $\ell$.
\end{defn}

In the context of QIT, it is often useful to extend certain operations to include extra resources~\cite{nielsen2002quantum}. Here, we consider adding ancillas (initialized in a product state) of identical Hilbert space $H_d$ to each lattice site. We then introduce the set of Local Unitaries (LU), denoted by ${\cal LU}$, as follows: $U\in {\cal LU}\subset {\cal U}$ if $U=\otimes_{i=1}^{M} u_i$, where $u_i$ acts only on the $i$-th local spin and its associated ancillas. We will consider ancillas and local unitaries as \emph{free resources}, i.e. we will be allowed to add as many ancillas as needed, and perform arbitrarily many local unitary operations.

When ancillas are available, we may modify the action of $V\in {\cal QC}_\ell$ by adding local operations between single layers of unitaries, that is $V^\prime =  U_{\ell}V_{\ell} \ldots U_2V_2 U_1V_1 U_0$, where $U_n\in {\cal LU}$. Note that, in general, $V^\prime$ is not a unitary operator on $H$, since $U_n$ also acts on the ancillas. Finally, we will consider an additional extension of the allowed operations, including LOCC: after a QC (which may include additional ancillas), we allow for local (orthogonal) measurements on the ancillas, and LU depending on the outcomes of the measurements, which are classically communicated among all the qudits. Classical communication will always be considered a \emph{free} operation.

\prlsection{State transformations with QC and LOCC}
The addition of measurements gives rise to randomness. Thus, by adding LOCC to QC, it might seem difficult to extend the class of states that can be prepared deterministically, but it is indeed possible. This is not surprising since in the context of QIT there are several instances where measurements, if followed by LOCC, can lead to deterministic transformations~\cite{gottesman1999demonstrating}, see e.g.~\cite{bennett1993teleporting,raussendorf2001one,verstraete2004valence}.

We address the question: when can a product state $|\mathbf{0}\rangle = |0\rangle^{\otimes M}\in H$ be (deterministically) transformed into another one, $\ket{\varphi}\in H$, using only QC or QC together with LOCC? For the first case, there exists $U\in {\cal QC}_{\ell}$, such that $|\varphi\rangle = U |\mathbf{0}\rangle$. For the latter, we restrict ourselves to the following scheme. We first apply a depth-$\ell$ circuit, with possibly local unitaries acting in between different layers of gates, as explained previously. Then, we sequentially measure each ancilla $a_i$ in some orthonormal basis, $\{|\varphi_{k_i}\rangle_i\}$, and apply $U\in{\cal LU}$ depending on the outcomes of all previous measurements (so, overall, we perform up to $M$ measurements and apply $M$ LU). Note that in this protocol we perform a single measurement per site. One could also define a more general scheme with multiple rounds of LOCC~\cite{bennett1996mixed}. While this would not change our conclusions, we restrict ourselves to the above definition.

\begin{defn}
[Transformations under QC and LOCC] We say that a state $\ket{\varphi}$ can be prepared by $X={\rm QC}_{\ell},{\rm QC}{\rm cc}_{\ell}$ if it can be obtained, respectively, by $U\in{\cal QC}_\ell$ or $U\in{\cal QC}_\ell$ together with LOCC, using the above procedures. We will write $|\mathbf{0}\rangle \xrightarrow{X} |\varphi\rangle$.
\end{defn}

Let us analyze the power of LOCC. For that we give a simple necessary condition for transformations using QC. In the following, we define the distance between two regions $A,B\subset \Lambda$ as $d(A,B)=\min_{i\in A, j\in B} d(i,j)$, where we denote by $d(i,j)$ the minimal number of edges connecting the vertices $i$ and $j$ in the graph associated with the lattice $\Lambda$.
\begin{prop}
\label{propQCA2}
Let $A,B\subset \Lambda$ with $d(A,B)>2\ell$ and $X_A$, $Y_B$ operators supported on $A$ and $B$ respectively. If $|\mathbf{0}\rangle \xrightarrow{{\rm QC}_\ell} |\varphi\rangle$, then
 \be\label{eq:necessary_condition}
 \langle \varphi|X_A Y_B|\varphi\rangle = \langle \varphi|X_A|\varphi\rangle\langle\varphi|Y_B|\varphi\rangle .
 \ee
\end{prop}
\noindent  See~\cite{Note2} for a proof. This proposition is useful to prove that some states cannot be prepared by QC, as we now exemplify.
\begin{example}[The GHZ and $W$ states] \label{Example1}
Let us consider qubits in a 1D lattice ($M=N$) with Periodic Boundary Conditions (PBC). The GHZ and $W$ states are~\cite{greenberger1989bell,dur2000three} 
 \be\label{eq:ghz_w}
 |{\rm GHZ}\rangle= \frac{1}{\sqrt{2}} (|0\rangle^{\otimes N} + |1\rangle^{\otimes N}),\ \ket{W}=\frac{1}{\sqrt{N}}\sum_{k=1}^N\sigma_k^-|0\rangle^{\otimes N}.
 \ee
For both states it is simple to find $X_A$, $Y_B$ with $d(A,B)=N/2$ s.t. \eqref{eq:necessary_condition} is not verified. Let us show that they can be prepared by ${\rm QC}{\rm cc}_{2}$. For $\ket{\rm GHZ}$, we attach one ancilla per site, except for the first one. We define a unitary acting on the $n$-th qubit and the $n+1$ ancilla as $u_n |0\rangle_{s_n}\otimes |0\rangle_{a_{n+1}} = |\Phi^+\rangle_{s_n,a_{n+1}}$ ($|\Phi^+\rangle_{s_n,a_{n+1}}$: maximally entangled Bell state) as well as $U=(\otimes_{n=1}^{N-1} u_n)\otimes v_N$, where $v=(\Id-i\sigma^y)/\sqrt{2}$. Applying $U$ to $|\mathbf{0}\rangle_{s,a}$ (which can be done with a QC of depth $2$), it generates  $\left(\otimes_{n=1}^{N-1} |\Phi^+\rangle_{s_n,a_{n+1}} \right)\otimes|+\rangle_{s_N}$ where $|+\rangle=(|0\rangle+|1\rangle)/\sqrt{2}$.
This state can be transformed into $\ket{\rm GHZ}$ via LOCC. To see this, we apply a local CNOT gate between each qubit and its ancilla, yielding $|\Phi\rangle= \sum_{\{k_n\}} |k_1\rangle_{s_1}
 \left(\otimes_{n=2}^N |k_n\rangle_{s_n} \otimes |k_{n-1}\oplus k_{n}\rangle_{a_n} \right)$, where $k_{n-1}\oplus k_{n}=k_{n-1}+ k_{n}$ (mod $2$), and measure all ancillas in the computational basis. Given the output $\{k_j\}_{j=2}^N$, we finally apply $\otimes_{n=2}^N (\sigma^x_n)^{\sum_{m=2}^n k_m }$  to the spins. With a similar construction, we can also prove $|\mathbf{0}\rangle \xrightarrow{{\rm QCcc}_2} |W\rangle$~\footnote{See Supplemental Material, which includes Refs.~\cite{arrighi2019overview,farrelly2020review, arrighi2011unitarity,farrelly2014causal,gross2012index,cirac2017matrix,sahinoglu2018matrix,duschatko2018tracking,gong2020topological,fidkowski2019interacting,piroli2021fermionic,piroli2020quantum,nielsen1999conditions,van2002renyi,bhatia2013matrix,schlingemann2001stabilizer,van2004graphical}, for details}.

\end{example}

We mention that related constructions for the GHZ state appeared before~\cite{meignant2019distributing}, see also~\cite{watts2019exponential}.

% ---------------------------------------------------------------
\begin{example}[Fixed points in $1D$]
\label{Example2}
In order to show the power of QCcc, we consider the fixed points of the Renormalization-Group (RG) procedure introduced in Ref.~\cite{verstraete2005renormalization}, representing a very general class of states in $1D$. To define them, we take a chain of $N$ sites with PBC, where each site is associated with three qudits $C_n$, $L_n$ and $R_n$ [center, left and right, respectively]. Up to LU transformations, RG fixed points take the form~\cite{cirac2017matrix_op}
\be\label{eq:rg_fp}
\ket{\Psi}=\sum_{k=1}^B \alpha_k \otimes_{n=1}^N\ket{k}_{C_n} \ket{\psi}_{R_{n},L_{n+1}}\,,
\ee
where $B\in \mathbb{N}$, $\alpha_k\in\mathbb{C}$, while $\ket{\psi}_{R_{n},L_{n+1}}$ is an entangled state between ${R_n}$ and $L_{n+1}$. Let us show that~\eqref{eq:rg_fp} can be prepared by ${\rm QC}{\rm cc}_{4}$. We introduce ancillas $C^\prime_n$, $L^{\prime}_n$, $R_n^\prime$, and create maximally entangled states between $R_n^\prime$ and $L_{n+1}$ with a depth-$2$ QC. Next, we prepare the qudits $C_n$ in the state $\sum_{k} \alpha_k \otimes_n\ket{k}_{C_n}$, which can be done by ${\rm QCcc}_2$, using ancillas $C_n^\prime$ and following the steps of Example~\ref{Example1}. Using LU, we then prepare the state $\ket{\psi}_{L^\prime_{n},R_{n}}$ between ancillas $L^\prime_{n}$ and $R_{n}$, conditioned to the state of $C_n$, i.e. $\ket{k}_{C_n}\ket{0}_{L^\prime_n}\ket{0}_{R_n}\mapsto \ket{k}_{C_n}\ket{\psi}_{L^\prime_{n},R_{n}}$. Finally, we use the entangled pairs between $R'_n$ and $L_{n+1}$ to teleport $L^\prime_n$ to $L_{n+1}$, which can be done via LOCC~\cite{bennett1993teleporting}.
\end{example}

\begin{example}[The Toric Code]
\label{Example3}

Finally, let us consider qubits in a 2D lattice with PBC ($M=N^2$), where $i\in \Lambda$ has two coordinates, $i=(i_1,i_2)$, and focus on the toric-code state, $|{\rm TC}\rangle$~\cite{kitaev2003fault}. For $N$ even, the TC can be defined by placing the qubits at the vertices of a square lattice. Let $P$ be the set of all plaquettes composed of four contiguous vertices forming a square. We divide them into two types, $P_A$ and $P_B$, following a chess-board pattern. For each A-plaquette $p\in P_A$, we introduce $X_p=\otimes_{i\in p} \sigma_{i}^x$, and define $|{\rm TC}\rangle \propto \prod_{p\in P_A} (\Id+X_p)
  |0\rangle^{\otimes M}$. We also set $S^\alpha_{j}=\otimes_{k=1}^N \sigma^\alpha_{j,k}$ for $j=1,\ldots,N$. %Since $S^z_{j}$ commutes with $X_p$, we have $S^z_{j} |{\rm TC}\rangle = |{\rm TC} \rangle$.
  It is well known that the TC can not be prepared by ${\rm QC}_\ell$ for $\ell$ independent of $N$~\cite{bravyi2006lieb}. This can also be seen by noticing that~\eqref{eq:necessary_condition} is not satisfied choosing $X_A=S^x_{1}$, $Y_B=S^x_{N/2+1}$. Let us show $|\mathbf{0}\rangle \xrightarrow{{\rm QCcc}_{16}} \ket{\rm TC}$. We do this following \cite{raussendorf2005long} (see also ~\cite{aguado2008creation}). For each $p\in P_A$, we include an ancilla, $a_p$ in the vertex at the upper-left corner of $p$. Next, we define the unitary $V=\prod_{p\in P_A}V_p$, with $V_p = \frac{1}{2} \left[ (\Id+X_p)\otimes \Id_{a_p} + (\Id-X_p)\otimes \sigma^x_{a_p}
 \right]$. $V_p$ may be implemented using $8$ nearest-neighbor gates: $(i)$ we introduce $4$ additional ancillas at the upper-left corner of $p$, denoted by $Q$; $(ii)$ we swap them with the qubits at the vertices of $p$ (with $4$ gates); $(iii)$ we apply (locally) $V_p$ to the five ancillas in $Q$; $(iv)$ we swap back the qubits in $Q$ with the vertices of $p$. Then, dividing $P_A$ into two subsets $P_A^{'}$, $P_A^{''}$ such that all plaquettes in each subset share no common qubit, we can implement $V$ by acting in parallel on all $p\in P_A^\prime$, then on all in $p\in P_A^{''}$, resulting in a QC of depth $16$. After applying $V$, we measure $\sigma^z$ in all the ancillas, $a_p$, with outcomes $k_p=\pm 1$. The fact that $\prod_{p\in P_A}X_p=\openone$ implies that the product of all $k_p$ equals one~\footnote{Indeed, if $\prod_{p\in P_A}k_p=-1$, using $\prod_{p\in P_A}X_p=\openone$ one can easily show that $\langle\psi_k|\psi_k\rangle=0$, where $\ket{\psi_k}$ is defined in~\eqref{eq:psi_k}.}. The resulting state is
 \be\label{eq:psi_k}
 |\psi_k\rangle \propto \prod_{p\in P_A} (\Id+k_p X_p)
  |0\rangle^{\otimes M}.
 \ee
Finally, it is easy to see that given a set of $k_p=\pm 1$ whose product equals one, it is always possible to find $Z_k$, a product of $\sigma^z$ operators, such that $ Z_k (\Id+k_p X_p) Z_k = (\Id+ X_p)$, $\forall p$. Thus, by applying the LU $Z_k$ we recover the TC deterministically. 
\end{example}

In summary, LOCC enlarge the set of states which can be prepared deterministically. One could wonder whether all states may be realized in this way. This is not the case, and only states satisfying an entanglement area law, similar to that characterizing ground states of local Hamiltonians~\cite{eiser2010area}, may be prepared. To see this, we have to consider a sequence of states $\{|\psi\rangle_M\}_M$ on lattices of increasing size. We assume that $|\psi\rangle_M$ is prepared by ${\rm QC}{\rm cc}_\ell$, where $\ell$ is independent of $M$, and denote by $S_0^{\psi}(A:A^c)$ the max-entropy between $A\subset V$ and its complement $A^c=V/A$, which upper bounds the von Neumann entropy~\cite{nielsen2002quantum}. We also call $\partial A$ the boundary of $A$, and denote by $|A|$ the number of qudits in $A$.

\begin{defn}[Entanglement Area Law] A sequence of states $\{|\psi_M\rangle\}_M$ obeys an entanglement area law if for all $A\subset V$, $S_0^{\psi_M}(A:A^c)\le c |\partial A|$, where $c$ is a constant independent of $M$.
\end{defn}

\begin{prop}
\label{prop:area_law}
Any sequence of states $\{|\psi_M\rangle\}_M$ prepared by ${\rm QCcc}_{\ell}$ (with $\ell$ independent of $M$) satisfies an entanglement area law.
\end{prop}
\noindent See~\cite{Note2} for a proof.

\prlsection{Phases of matter} QC appear naturally in the standard classification of topological phases of matter~\cite{chen2010local,hastings2013classifying,chiu2016classification,zeng2015quantum,zeng2015gapped}. Colloquially, for ground states of gapped, local Hamiltonians, it is known that if two states are in the same phase (i.e. their parent Hamiltonians are connected by a differentiable path of gapped, local Hamiltonians), then they are mapped onto one another by a ``low-depth'' QC. Inverting the logic, one could use QC to define equivalence classes. However, some care must be taken: indeed if $\ket{\psi_2}=U\ket{\psi_1}$, and $\ket{\psi_3}=V\ket{\psi_2}$ with $U,V\in {\cal QC}_\ell$, then to transform $\ket{\psi_1}$ to $\ket{\psi_3}$ may require an operation in ${\cal QC}_{2\ell}$, meaning that one has to allow for the depth to change. One way to do this is to define an equivalence relation between state sequences, $\Psi=\{|\psi_M\rangle\in H_M\}_{M=M_0}^\infty$, for lattices of increasing size, where $M_0\in\mathbb{N}$: one can say that $\Psi\sim \Phi$ if $\exists U_M\in {\cal QC}_{f(M)}$ s.t. $||\ket{\psi_M}- U_M\ket{\varphi_M}||\xrightarrow{M\to\infty} 0$. Here, $f(M)$ is a function that grows sufficiently slow in $M$. For example, ground states of gapped, local Hamiltonians in the same phase are equivalent by this definition choosing $f(M)$ to be a polylogarithmic function of $M$~\cite{coser2019classification,haah2021quantum} (where one also allows for a number of ancillas polylogarithmic in $M$), see also Refs.~\cite{osborne2006efficient,osborne2007simulating, bachmann2012automorphic,huang2015quantum}.

We wish to extend this definition by replacing ${\rm QC}$ with ${\rm QC}{\rm cc}$ (and without restricting to ground states). To do that, we allow for approximate preparation protocols, where a pure state may be mapped onto a mixed state $\rho$, as we now explain~\footnote{We note that other definitions which do not modify our conclusions are possible}. A given preparation protocol in QCcc$_{\ell}$ (where ancillas are traced out at the end), defines a quantum channel ${\cal C}$~\cite{nielsen2002quantum}. If a pure initial state $\ket{\varphi}$ can be mapped onto the (mixed) state $\sigma$ for some ${\cal C}$ defined in this way, we will write $\ket{\varphi}\xrightarrow{{\rm QCcc}_{\ell}} \sigma$. We will also use the symbol QCcc$^{(k)}_{\ell}$ to denote transformations obtained by composing $k$ such channels $\{{\cal C}_j\}_{j=1}^k$. Then, we may define an equivalence relation as follows. First, given two sequences $\Psi$, $\Phi$, we write $\Psi\mapsto \Phi$ if $\exists k\in \mathbb{N}$ and a sequence of (mixed) states $\{\sigma_M \}_{M=M_0}^\infty$, s.t. $\ket{\psi_M} \xrightarrow{{\rm QCcc}^{(k)}_{f(M)}} \sigma_M$ and $||\sigma_M-\ket{\varphi_M}\bra{\varphi_M}||_1\xrightarrow{M\to\infty}  0$, where $||\cdot||_1$ is the trace norm. Here, analogously to the case of QC, we choose $f(M)$ to be a polylogarithmic function of $M$. Finally, we say that $\Psi$ is QCcc-equivalent to $\Phi$, if $\Psi\mapsto \Phi$ and $\Phi\mapsto \Psi$. Note that this more complicated definition is needed to ensure symmetry and transitivity (which simply follows from contractivity of the trace norm).

Solving the full classification problem is expected to be very hard. However, based on Example~\ref{Example2}, we can give a strong result in $1D$, proving that all translational invariant MPS with fixed bond dimension belong to the trivial class~\cite{Note2}.
\begin{thm}
\label{MPS_classification}
In $1D$, all translational invariant MPS with fixed bond dimension are in the same phase as the trivial state.
\end{thm}
\noindent A proof of this theorem is given in~\cite{Note2}. QCcc classes are strictly larger than those in the standard classification of topological phases, as exemplified by the TC. In fact, it is natural to conjecture that the same is true for all non-chiral topologically-ordered states, although this problem goes beyond the scope of this work. The suggested classification is expected to have practical ramifications in preparation protocols with NISQ devices, as states in the trivial phase may be prepared efficiently with operations already at hand -- geometrically local gates and on-site measurements. At  the same  time, from the fundamental standpoint, QCcc provide a unified framework where the locality-based classification of states in QMBP and QIT meet.

\prlsection{Unitary operations}
It is known that allowing for post-selection processes the power of quantum computers increases~\cite{aaronson2005quantum,schuch2007computational}. Post-selection, however, has practical limitations, due to the large number of times that a computation must be performed. Here we take a different point of view and ask whether, combining LOCC and QC, one can implement \emph{deterministically} a larger set of unitary operations beyond QC~\cite{gottesman1999demonstrating}. This is different from the state-transformation protocols, since now we want unitary actions on all possible input states.

Let us now introduce a general scheme to implement unitary operators, which involves QC and LOCC~\footnote{Note that, even without measurements, i.e. by simply allowing for additional ancillas, the set of unitary operations is enlarged~\cite{Note2}}. First, we prepare a state $\ket{\phi}_a$ on the ancillas, using only QC and LOCC as in the state-transformation protocol discussed before. Then, given an input state $\ket{\psi}$, the procedure consists in applying a depth-$\ell$ quantum circuit $V$ (including ancillas and LU) to the pair system-ancilla, followed by LOCC. In particular, we consider operations
\be
\label{unitary_action}
\ket{\psi}\to U_s^{\alpha}(\otimes_k\bra{\alpha_k}) V_{sa} (\ket{\psi}_s\otimes |\phi\rangle_a)\,.
\ee
The subscripts $s$ and $a$ label system and ancilla, $\ket{\alpha_k}$ is an element of a local orthonormal basis for the ancillas, while $U^{\alpha}_s\in {\cal LU}$, which might depend on the outcomes $\alpha_k$. We are interested in the special cases where the action~\eqref{unitary_action} defines a unitary operation.

\begin{defn}[LOCC-assisted quantum circuits] ${\cal QC}{\rm cc}_{\ell}\subset {\cal U}$ is the set unitary operators that can be implemented (deterministically) by a QC of depth $\ell$ with the help of ancillas using the protocol~\eqref{unitary_action}.
\end{defn}
\noindent Note that in the above we also require $|\mathbf{0}\rangle \xrightarrow{{\rm QCcc}_\ell} |\phi\rangle$.

Trivially, ${\cal QC}_{\ell}\subseteq {\cal QC}{\rm cc}_{\ell}$. In fact, the inclusion is strict, as we illustrate with a specific construction which, on the one hand, ensures that the map~\eqref{unitary_action} is unitary, while, on the other, allows us to implement operators beyond QC. Before that, we need to recall two notions in QIT. The first one is that of Clifford operators~\cite{gottesman1997stabilizer,gottesman1998theory}. To define them, we introduce the set ${\cal Q}$ of tensor products of Pauli operators, i.e.  ${\cal Q} = \{ \otimes_{i=1}^M \sigma^{\alpha_i}_i, \quad \alpha_i=0,x,y,z\}$, where $\sigma^0_j=\openone_j$. Then, $U\in {\cal U}$ is a Clifford operator if for any $s\in {\cal Q}$, $U^\dagger s U=s'\in {\cal Q}$ (possibly up to a factor). The second one, is that of Locally Maximally-Entanglable (LME) states~\cite{kraus2009local}. They are defined as the states $\ket{\varphi}_s$ for which there exist LU which create a maximally entangled state between the spins and the ancillas, i.e. there exist $u_i\in {\rm LU}$ such that $|R\rangle =\otimes_{i=1}^M u_i (|\varphi\rangle_s \otimes |\mathbf{0}\rangle_a)$ fulfills ${\rm tr}_a(|R\rangle\langle R|)= \Id_s$.

Now, our construction is as follows. First, we append one ancilla per site, $a_n$, and prepare $\ket{\phi}_a$ by QCcc$_\ell$. We require that $\ket{\phi}$ is a stabilizer state, i.e. the unique common eigenstate of a set of commuting elements of ${\cal Q}$. This implies that $\ket{\phi}$ is also LME~\cite{kraus2009local}. Thus, adding one additional ancilla per site, $a^\prime_n$, there exists $V\in {\cal LU}$ s.t $\ket{R}_{a,a^\prime}=V\ket{\phi}_a\ket{0}_{a^\prime}$ is maximally entangled. This implies that the map $|\psi\rangle_s \mapsto d^{M} \;_{s,a^{\prime}}\langle \Phi^+ |R\rangle_{a,a^\prime}\otimes |\psi\rangle_{s}$ equals the action of a unitary operator $U$~\cite{wolf2012quantum} (where $\ket{\Phi^+}_{s,a}$ is the maximally entangled Bell state between system $s$ and ancillas $a$). Furthermore, since $\ket{\phi}$ is a stabilizer, one can choose $V$ such that $U$ is a Clifford operator~\cite{Note2}. Then, for any input $\ket{\psi}$, $U\ket{\psi}$ can be implemented deterministically using LOCC. To do this, we perform a Bell measurement on the qubits $s_n$ and $a^\prime_n$. This produces $U(\otimes_n\sigma^{\alpha_n})\ket{\psi}_{a}$, where $\alpha_n$ depend on the values of the measurement. Since $U$ is a Clifford operator, $U(\otimes_n\sigma^{\alpha_n})=wU$, with $w\in {\cal Q}$ and hence $w\in {\cal LU}$, so that $U\ket{\psi}$ is recovered applying $w^\dagger \in {\cal LU}$.

\begin{example}[The GHZ and TC unitaries]
Consider the GHZ state. It is a stabilizer state prepared by QCcc$_2$, and thus it may be used to implement a unitary operator. To see this explicitly, starting from $\ket{{\rm GHZ}}_s$,  we first apply a single phase gate to one of the state qubits. Then, we prepare a maximally entangled state with an ancillary system where all ancillas are initialized in $\ket{+}$, by simultaneously applying CNOT gates to each system-ancilla pair (with the system being the target), thus obtaining a state $\ket{R}_{s,a}$. It is easy to see that the action $|\psi\rangle_{a^{'}} \mapsto d^{M} \;_{a,a^{'}}\langle \Phi^+ |R\rangle_{s,a}\otimes |\psi\rangle_{a^{'}}$ corresponds to the unitary $U_{{\rm GHZ}} = \left( \Id + i \sigma_x^{\otimes M} \right)/{\sqrt{2}}$. Importantly $U_{{\rm GHZ}}$ is a Clifford operator, and thus may be implemented by LOCC. Also, $U_{\rm GHZ}\notin {\cal QC}_\ell$ for $\ell<N$, because $U^\dagger \sigma^z_1 U$ is a string of Pauli matrices over the whole system. Similarly, starting from the TC, we can construct a unitary $U_{\rm TC}\in {\cal QC}{\rm cc}_{16}$ s.t. $U_{\rm TC}\ket{{\bf 0}}$ is locally equivalent to $\ket{{\rm TC}}$~\cite{Note2}, implying that $U_{\rm TC}\notin {\cal QC}_{\ell}$ for $\ell<N/4$.
\end{example}
% -----------------------------------------------------

\prlsection{Outlook} In this Letter we have introduced a paradigm to  classify states and operations based on a notion  of locality inspired by both QIT  and QMBP, arguing for its fundamental relevance and potential practical  importance. Our work raises several questions. First, we have seen examples of topologically-ordered states in the trivial class, but an obvious question is  whether all representatives of non-trivial phases may be prepared by QCcc. A similar problem holds for chiral states, which we have not addressed. Moreover, we have considered here QC composed  of local gates; it is natural to wonder how our conclusions are modified using non-local gates instead. Finally, ideas related to those presented here may lead to a classification for unitary operators: although this requires to get around some subtleties, we expect that such a classification will be different from the one for the corresponding Choi-Jamiolkowski states~\cite{nielsen2002quantum}. We leave these questions for future work.

\prlsection{Acknowledgments} We thank Alex Turzillo for very useful discussions. We acknowledge support by the EU Horizon 2020 program through the ERC Advanced Grant QUENOCOBA No. 742102 and by the DFG (German Research Foundation) under Germany’s Excellence Strategy --  EXC-2111 -- 390814868. This project has received funding from the European Union’s Horizon 2020 research and innovation programme under grant agreement No 899354.

% ---------------------------------------------------------------
% ---------------------------------------------------------------

\bibliography{./bibliography}

\clearpage
\newpage

\appendix

\section*{\large Supplemental material}

Here we will provide additional details about the results stated in the main text.

\section{Quantum Cellular Automata}
\label{sec:QCA}

Some of the statements presented in the main text are naturally proven using the notion of Quantum Cellular Automata (QCA), which we introduce in this section.

The set of QC can be extended to a larger class of unitaries, by simply allowing for additional ancillas and LU. Specifically, let us consider $V^\prime =  U_{\ell}V_{\ell} \ldots U_2V_2 U_1V_1 U_0$, where $V_{n}$ are layers of quantum gates acting on disjoint pairs of nearest-neighbor spins, while $U_n\in {\cal LU}$. We recall that $V_n$ acts only on the physical qudits, while $U_n$ acts on the local qudit $n$ and all its associated ancillas. Suppose there exist $W,\tilde{W}\in {\cal U}$ such that for all $\ket{\psi}\in H$ we have $V^\prime \left( |\psi\rangle_s\otimes|\mathbf{0}\rangle_a \right) = W|\psi\rangle_s \otimes \tilde{W}|\mathbf{0}\rangle_a$. We claim that, in general, $W$ is not a QC whose depth is independent of the system size. A simple example, which we will detail later, is given by the shift-operator. In general, the unitaries constructed in this way are QCA~\cite{arrighi2019overview,farrelly2020review}, which are known to strictly contain QC.

In order to define QCA, we need to introduce some notation. We denote by $d(i,j)$ the distance between two lattice sites, $i,j\in\Lambda$, as the minimal number of edges that connects them. We also define the distance between two sets of sites, $A,B\subset \Lambda$ as $d(A,B) = \min d(i,j)$ where the minimum is taken over all $i\in A,j\in B$. Given an operator $X$ acting on the spins, we define its support as the minimal subset $A=\{i\in \Lambda\}$  for which  $X=\Id_{A^c}\otimes {\rm tr}_{A^c}(X)/d^{|A^c|}$ ($A^c=\Lambda\setminus A$ is the complement of $A$, while $|A|$ denotes the number of sites in $A$), i.e, where it acts non-trivially. From now on, we will write $X_A$ for an operator $X$ supported on (or within) $A$, and $X_i$ if it is supported on a single site, $i\in \Lambda$. A unitary operator $U$ defines a map (in the Heisenberg picture) between operators that, in general, will change their support: $U^\dagger X_i U= \tilde X_{\bar i}$, where $\bar i \subseteq \Lambda$. We define the range of $U\in{\cal U}$,
\be
r_U = \max [\max_{j\in \bar{i}}d(i,j)]\,,
\ee
where the outer maximization is with respect to $i\in \Lambda$ and all operators $X_i$ with support on $i$. With this definition, for any $A\subset \Lambda$ and $\tilde X_{\bar A}=U^\dagger X_A U$, we have that $d(A,\bar A)\le r_U$, where we denoted by $\bar{A}$ the support of the transformed operator.

\begin{defn}
[Range-$r$ Quantum Cellular Automata] ${\rm QCA}_{r}\subset {\cal U}$ is the set of unitary operators whose range is at most $r$.
\end{defn}

There is a very close connection between QCA and QC. On the one hand, it is trivial to see that ${\cal QC}_\ell\subset$QCA$_{\ell}$. On the other hand, it can be proven that any QCA with finite range $r$, may be implemented by a QC $V^\prime =  U_{\ell}V_{\ell} \ldots U_2V_2 U_1V_1 U_0$ (thus also acting on the ancillas) of depth $\ell$ which only depends on $r$ and $D$, but not on the system size~\cite{arrighi2011unitarity} (see also Refs.~\cite{farrelly2014causal,farrelly2020review}). In general, however, if ancillas are not available, then it is not possible to represent QCA of finite range by QC of depth independent of $N$. As a simple example, let us take the left-shift operator, $T$, defined by
 \be
 T|n_1,\ldots,n_N\rangle = |n_2,\ldots,n_1\rangle,
 \ee
 which is clearly a QCA with range $1$. First, if ancillas are available, then it is easy to show that $T$ can be implemented by a QC of depth $2$. To see this, we append one ancilla per site, and consider the unitary $W=T\otimes T^{\dagger}$, acting on the doubled Hilbert space $H\otimes H$. It is immediate to see that $W=S^\prime S$, where $S=\otimes_j S_{s_j,{a_{j-1}}}$, and $S^\prime = \otimes_j S_{s_j,{a_{j}}}$, where $S_{s_i,a_j}$ swaps the qubit $s_i$ and the ancilla $a_j$. On the other hand, $T\notin {\cal QC}_\ell$ with $\ell< N/2$. This can be seen as a simple application of the index theory for QCA, first introduced in ~\cite{gross2012index} for infinite systems, and later studied in Refs.~\cite{cirac2017matrix,sahinoglu2018matrix,duschatko2018tracking,gong2020topological} for finite sizes (see also Refs.~\cite{fidkowski2019interacting,piroli2021fermionic} for an extension to fermionic systems).

\section{Proof of Prop.~\ref{propQCA2}}

The proof of Prop.~\ref{propQCA2} follows immediately from the general results for QCA of Ref.~\cite{piroli2020quantum}, by noticing that, if $U\in {\cal QC}_{\ell}$, then $U$ is a QCA of range $r_U\leq \ell$.

\section{The W-state}
\label{sec:w_state}

In this section, we show that $|\mathbf{0}\rangle \xrightarrow{{\rm QCcc}_2} |W\rangle$. We consider an array of $N$ qubits $\{s_n\}_{n=1}^N$ in $1D$. For $n=1,2,\ldots N-1$, we take two ancillas per site, denoted by $a_{n,l}$, $a_{n,r}$, while for the last site we take three, denoted by $a_{N,l}$, $a_{N,r}$ and $a_{N+1,l}$, respectively. We assume that all the qubits and ancillas are initialized in the state $\ket{0}$. Let us define
\be
\ket{\psi}=\otimes_n \ket{0}_{s_n}\otimes\ket{0}_{a_{1,l}}\otimes_{n=1}^{N}\ket{\Phi^{+}}_{a_{n,r},a_{n+1,l}}\,.
\ee
Here $\ket{\Phi^{+}}_{a_{n,r},a_{n+1,l}}$ is the maximally entangled Bell state between ancillas $a_{n,r}$ and $a_{n+1,l}$. It is immediate to show that $\ket{\psi}$ can be created by a QC of depth $2$ (including LU acting also on the ancillas). Let us show that $\ket{W}$ can be created from $\ket{\psi}$ using LOCC. To this end, we apply a LU to $s_1$, and map it to $\ket{s_1}=x_1\ket{1}+y_1\ket{0}$, where $x_1^2+y_1^2=1$, and where $x_1\in \mathbb{R}$ will be defined later. For $z\in\mathbb{R}$, we also define a two-qubit unitary operator $V_{1,2}(z)$ s.t.
\begin{align}
V_{1,2}(z)\ket{0}_1\ket{0}_2&=\ket{0}_1\ket{0}_2\,,\\
V_{1,2}(z)\ket{0}_1\ket{1}_2&=z\ket{0}_1\ket{1}_2+\sqrt{1-z^2}\ket{1}_1\ket{0}_2\,.
\end{align}
Next, we apply $V_{a_{1,l},s_{1}}(z_1)$ to the sites $a_{1,l}$ and $s_1$, with $z_1=1/(x_1\sqrt{N})$. Then we use the entangled state between $a_{1,r}$ and $a_{2,l}$ to teleport the state in $a_{1,l}$ to $a_{2,l}$ (which can be done with local measurements and LU). As a result, the state of the spins $s_{1}$ and $a_{2,l}$ is
\be
\left(y_1\ket{0}_{s_1}+\frac{1}{\sqrt{N}}\ket{1}_{s_1}\right)\ket{0}_{a_{2,l}}+x_2\ket{0}_{s_1}\ket{1}_{a_{2,l}}\,,
\ee
where $x_2=\sqrt{x^2_1-\frac{1}{N}}$. Now, we perform a swap between $a_{2,l}$ and $s_2$ and repeat this procedure starting at site $2$, changing the argument of $V(z)$. In particular, we apply $V_{a_{2,l},s_{2}}(z_2)$ to the sites $a_{2,l}$ and $s_2$, with $z_2=1/(x_2\sqrt{N})$. Then we use the entangled state between $a_{2,r}$ and $a_{3,l}$ to teleport the state in $a_{2,l}$ to $a_{3,l}$. As a result, the state of the qubits $s_{1}$, $s_2$ and $a_{3,l}$ is
\begin{align}
\left(y_1\ket{0}_{s_1}\ket{0}_{s_2}+\frac{1}{\sqrt{N}}\ket{1}_{s_1}\ket{0}_{s_2}+\frac{1}{\sqrt{N}}\ket{0}_{s_1}\ket{1}_{s_2}\right)\ket{0}_{a_{3,l}}\nonumber\\
+x_3\ket{0}_{s_1}\ket{0}_{s_2}\ket{1}_{a_{3,l}}\,,
\end{align}
where $x_{3}=x_{2}\sqrt{1-z_2^2}=\sqrt{x_1^2-2/N}$.
We iterate this procedure, choosing at the $n$-th step $z_n=1/(x_n\sqrt{N})$. At the last step of the iteration, corresponding to site $N$, the state of the qubits $s_{1}$,..., $s_N$ and $a_{N+1,l}$ is
\begin{align}
\left(y_1\ket{0}_{s_1}\otimes \cdots \otimes \ket{0}_{s_N}+\ket{W}\right)\ket{0}_{a_{N+1,l}}\nonumber\\
+x_{N+1} \ket{0}_{s_1}\otimes \cdots \otimes \ket{0}_{s_N}\ket{1}_{a_{N+1,l}}\,,
\end{align}
with $x_{N+1}=\sqrt{x_1^2-1}$. Choosing $x_1=1$, we have $y_1=0$, $x_{N+1}=0$, and the state of the qubits $s_{1}$,...$s_N$ factorizes, becoming equal to $\ket{W}$.

\section{Proof of Prop.~\ref{prop:area_law}}

Consider the region $A\subset \Lambda$. We denote by $A^{\prime}$ the set of ancillas associated with region $A$, and with $A^c$ the complement of A. Suppose $|\mathbf{0}\rangle_M \xrightarrow{{\rm QCcc}_{\ell}} \ket{\psi}_M$. This means that we can obtain $\ket{\psi}_M$ by first acting on $|\mathbf{0}\rangle_M$ with $V^\prime =  U_{\ell}V_{\ell} \ldots U_2V_2 U_1V_1 U_0$, where $U_n\in {\cal LU}$ and then applying LOCC. After applying each layer $V_n$, since $V_n$ only acts on the physical qudits, the entanglement $S_0(AA^{\prime}:(AA^{\prime})^c)$ increases at most by $c|\partial A|$, where $c$ is a constant that only depends on the local dimension $d$~\cite{piroli2020quantum} (so it does not depend on the number of ancillas per site). On the other hand, $U_n\in {\cal LU}$, so it does not increase $S_0(AA^{\prime}:(AA^{\prime})^c)$. This is also true for LOCC, which do not increase the bipartite entanglement~\cite{nielsen1999conditions,van2002renyi}. Finally, we note that, by definition of QCcc, after LOCC the final state is factorized wrt to the bipartition system-ancillas, i.e. $\rho_{AA^\prime}=\rho_A\otimes \rho_{A^\prime}$, where $\rho_A$ is the density matrix reduced to the region $A$. Using additivity of entanglement, Prop.~\ref{prop:area_law} then immediately follows.

\section{Proof of Theorem~\ref{MPS_classification}}

In this section, we present the proof of Theorem~\ref{MPS_classification}.  
We will focus on translational invariant MPS 
\be
\left|\phi_{N}\right\rangle=\sum_{s_{1}, \ldots, s_{N}} \operatorname{tr}\left(M^{s_{1}} \ldots M^{s_{N}}\right)\left|s_{1}, \ldots, s_{N}\right\rangle\,,
\ee
where $M^s$ are $\chi\times \chi$ matrices, and $\chi$ is called the bond-dimension.  Any such MPS can be brought into a canonical form~\cite{cirac2017matrix_op}, that is,
\be
M^{i}=\bigoplus_{k=1}^{r} \mu_{k} M_{k}^{i}\,,
\ee
and $M^i$ are \emph{normal} tensors. This means that: $(i)$ there exists no non-trivial projector $P$ such that $M^iP = PM^iP$; $(ii)$ its associated transfer matrix, has a unique eigenvalue of magnitude (and value) equal to its spectral radius, which is equal to one. Hence, in the following we can assume without loss of generality that $\left|\phi_{N}\right\rangle$ is in canonical form. Furthermore, we recall that, for any integer $q$, we can construct a new MPS $\left|\phi^q_{N}\right\rangle$ on a chain of $N/q$ qudits of local dimension $d^q$, by grouping together blocks of $q$ neighboring sites (i.e. \emph{blocking} $q$ times). The proof consists of two parts. First, we show that $\left|\phi^q_{N}\right\rangle$ can be approximated, up to an error $\varepsilon=O(Ne^{-\beta q})$ for some $\beta>0$, by an MPS $\ket{\tilde{\phi}^q_N}$ which is a fixed point for the RG procedure introduced in Ref.~\cite{verstraete2005renormalization}. Second, we prove that such a fixed point can be prepared by QCcc$_{f(q)}$,  where $f(q)$ is a polynomial function of $q$.  From these two facts, Theorem~\ref{MPS_classification} easily follows. 

Let us first consider the RG fixed point $\ket{\tilde{\phi}^q_N}$ and prove the second part. As we have already mentioned, RG fixed points are locally equivalent to the states~\eqref{eq:rg_fp}~\cite{cirac2017matrix_op}. However, here one needs to be careful about the notion of locality: since $\ket{\tilde{\phi}^q_N}$ is obtained by blocking $q$ qudits, a LU in the blocked lattice corresponds to an operator $U\in \mathcal{U}$ acting on a set $A_q\subset \Lambda$ of $q$ adjacent qudits in the unblocked chain. Still, we can implement $U$ with the following procedure: $(i)$ we swap all the qudits in $A_q$ with ancillary ones, associated with a single qudit $s_k\in A_q$; $(ii)$ we apply $U$ locally, on the qudit $s_k$ and all its ancillas; $(iii)$ we swap back the ancillas with the qudits in $A_q$. This allows us to implement $U$ on $A_q$ with a sequence of $n<q^2$ nearest-neighbor gates. Since this procedure can be done in parallel for all the $N/q$ blocks, we can transform the state $\ket{\tilde{\phi}^q_N}$ into a state $\ket{\chi^q_N}$ of the form~\eqref{eq:rg_fp} with a QC of depth $n<q^2$. Finally, we need to show that $\ket{\chi^q_N}$ can be prepared by QCcc$_{f(q)}$, with $f(q)$ a polynomial function of $q$. To do that, we can follow the procedure of Example~\ref{Example2}, where, again, one needs to be careful about the notion of locality. Repeating the argument of before, any one- and two-site gate in the blocked chain may be performed using $n<2q^2$ sequential two-site gates in the original lattice. Note that, in the construction of Example~\ref{Example2}, all local operations can be performed in parallel, so that their action on the original lattice may be performed as a quantum circuit of depth polynomial in $q$. Similarly, one has to be careful when performing measurements. Indeed, in order to repeat the construction of Example~\ref{Example2}, one needs to take joint measurements on qudits that are not at the same site with respect to the unblocked lattice. This can be done by first moving all the qudits into the ancillary space of a single qudit with swaps, and then performing the joint measurement in the associated local space. Again, it is important that all the measurements in Example~\ref{Example2} can be performed in parallel. Putting all together, we find that $\ket{\chi^q_N}$ can be prepared by QCcc$_{f(q)}$, where $f(q)$ is a polynomial function of $q$.

Let us now prove the second part, which requires a more technical analysis. In what follows we will denote by $||\cdot ||_{\infty}$,  $||\cdot ||_{1}$ and $||\cdot ||_{F}$ the operator norm, the trace norm, and the Frobenius (or Hilbert-Schmidt) norm, respectively. For simplicity, we will assume that $M^s$ is normal. The proof for the general case is completely analogous, although it requires more cumbersome notation.  Since $M^s$ is normal, the transfer matrix $\tau$ has a unique largest eigenvalue $\lambda_0=1$, associated with a trivial Jordan block~\cite{cirac2017matrix_op}. Note that $\left|\phi_{N}\right\rangle$ is not normalized at finite $N$, but $||\left|\phi_{N}\right\rangle||\to 1$ for $N\to\infty$ .  Let $\lambda_k$ be the other eigenvalues of $\tau$ with $|\lambda_k|<1$ and associated Jordan block $J_{r_k}(\lambda_k)$, with dimension $r_k<\chi^2$. We call $|\lambda_1|=e^{-\alpha}$ the absolute value of the second largest eigenvalue of $\tau$. By blocking $q$ times, we obtain a new MPS on a chain of length $M=N/q$, which we denote by
\be
\left|\varphi_{M}\right\rangle=\sum_{s_{1}, \ldots, s_{M}} \operatorname{tr}\left(A^{s_{1}} \ldots A^{s_{M}}\right)\left|s_{1}, \ldots, s_{M}\right\rangle\,.
\ee
We introduce the graphical notation
\be\label{eq:a_tensor}
A=
\begin{tikzpicture}[baseline={([yshift=-2.5ex]current bounding box.center)}, scale=0.7]
	
	\draw (0,0)  -- (0,0.7) node[above]{};
	\draw (-0.7,0) node[left]{} -- (0.7,0) node[right]{};	
	\filldraw[fill=white,draw=black] (0,0) circle [radius=0.25];
\end{tikzpicture}\,,
\ee
and also the transfer matrix
\bea\
\tau_{AA}=
\begin{tikzpicture}[baseline={([yshift=-0.5ex]current bounding box.center)}, scale=0.7]
	
	\draw (0,-0.25) -- (0,1) node[left]{};
	\draw (-0.5,0) node[left]{} -- (0.5,0) node[right]{};
	\draw (-0.5,1) node[left]{} -- (0.5,1) node[right]{};
	
	\filldraw[fill=white,draw=black] (0,0) circle [radius=0.25];
	\filldraw[fill=white,draw=black] (0,1) circle [radius=0.25];
	
\end{tikzpicture}\,.
\label{eq:transfer_m}
\eea
By construction $\tau_{AA}$ has a largest eigenvalue $\lambda_0=1$, while the second largest one satisfies $|\lambda_1|=e^{-q\alpha}$. Then
\be
\tau_{AA}=\tau_{BB}+R\,.
\ee
Here $R$ contains all the Jordan blocks associated with the subleading eigenvalues of $R$,  while $\tau_{BB}=\ket{a}\bra{b}$ and $\tau_{BB}^2=\tau_{BB}$, where $\ket{a}$, $\bra{b}$ are the right and left fixed points of $\tau_{BB}$, respectively. Using that 
\be
||J_{r_k}(\lambda_k)^q||_{\infty}\leq \Gamma(q) e^{-\alpha q}:=\chi^3  q^{\chi^2-1} e^{\alpha (\chi^2-1)} e^{-\alpha q}
\ee
and the fact that $R=V^{-1} (\oplus_k J_{r_k}^q(\lambda_k )V$, where $V$ is a fixed gauge transformation, we obtain $||R||_{\infty}\leq C_V \Gamma(q) e^{-\alpha q}$, with $C_V=||V^{-1}||_{\infty}||V||_{\infty}$ and so
\be\label{eq:difference}
||R||_F\leq \Lambda(q) e^{-q\alpha}\,.
\ee
where $\Lambda(q)= \chi  C_V \Gamma(q)$. It is useful to use the Frobenius norm, because it does not depend on which indices are used as input and which indices are used as output (whereas the trace norm does).  

Let us interpret now $\tau_{AA}$ as a matrix where the input (output) indices are the lower (upper) lines in the graphical representation~\eqref{eq:transfer_m}. With this choice, we have $\tau_{AA}=A^\dagger A$, where $A$ is interpreted as a $d\times \chi^2$ matrix, with input (output) indices defined by the lower (upper) lines of the graphical representation~\eqref{eq:a_tensor}. Let us define $\tilde{A}=\sqrt{\tau_{AA}}=\sqrt{A^\dagger A}$, using the standard definition for the square root of an Hermitian matrix~\cite{bhatia2013matrix}. It is immediate to show that $A=U\tilde{A}$, where $U$ is a $d\times \chi^2$ isometry, satisfying $U^\dagger U=\openone_{\chi^2}$ . We also define $\tilde{B}=\sqrt{\tau_{BB}}$,
 $B=U \tilde{B}$ and
\be
\left|\psi_{M}\right\rangle=\sum_{s_{1}, \ldots, s_{M}} \operatorname{tr}\left(B^{s_{1}} \ldots B^{s_{M}}\right)\left|s_{1}, \ldots, s_{M}\right\rangle\,.
\ee
Note that $\ket{\psi_M}$ is an RG fixed point by construction, because $B^\dagger B=\tau_{BB}$, and $\tau_{BB}^2=\tau_{BB}$~\cite{cirac2017matrix_op} [in the last equality, $\tau_{BB}$ is indented to be a matrix with input (output) associated with the right (left) legs of the grafical representation~\eqref{eq:transfer_m}]. Accordingly
\be
\braket{\psi_M|\psi_M}={\rm tr}[\tau_{BB}^M]=1\,.
\ee

We want to show that $\ket{\psi_M}$ is close to $\ket{\varphi_M}$. To this end, we compute
\be
|\braket{\psi_M|\varphi_M}-1|=|{\rm }{\rm tr}[\tau_{AB}^M-\tau_{BB}^M]|\,.
\ee
Here we defined $\tau_{AB}=A^\dagger B$, where, as before, $A$ and $B$ are interpreted as $d\times \chi^2$ matrices. We have
\begin{align}
\ &|{\rm }{\rm tr}[\tau_{AB}^M-\tau_{BB}^M]|\leq  ||\tau_{AB}^M-\tau_{BB}^M||_1\nonumber\\
=&||\sum_{k=0}^{M-1}\tau_{AB}^{M-1-k}(\tau_{AB}-\tau_{BB})\tau_{BB}^k||_1\nonumber\\
\leq& \sum_{k=0}^{M-1}||\tau_{AB}^{M-1-k}||_{\infty}{\rm max}(1,||\tau_{BB}||_{\infty})||(\tau_{AB}-\tau_{BB})||_{1}\label{eq:final_eq}
\end{align}
where we used $\tau_{BB}^k=\tau_{BB}$ for $k\neq 0$, and that, for all matrices $X$, $Y$, $||XY||_{1}\leq ||X||_{\infty} ||Y||_1$. Next, we use $||(\tau_{AB}-\tau_{BB})||_{1}\leq \chi ||(\tau_{AB}-\tau_{BB})||_{F}$. The last expression is in terms of the Frobenius norm, which does not depend on which indices of the matrices $\tau_{AB}$ and $\tau_{BB}$ are interpreted as input and output. Thus, as before, we can choose  the input to be the lower lines, and the output to be the upper lines. With this choice, $\tau_{AB}=B^{\dagger}A={\tilde{B}}^{\dagger}\tilde{A}$, while $\tau_{BB}={\tilde{B}}^\dagger \tilde{B}$, and
\begin{widetext}
\begin{align}\label{eq:intermediate}
||(\tau_{AB}-\tau_{BB})||_F&=||\tilde{B}^\dagger(\tilde{A}-\tilde{B})||_F \leq  \chi||\tilde{B}^\dagger||_{\infty}||(\tilde{A}-\tilde{B})||_{\infty} \nonumber\\
&\leq  \chi||\tilde{B}||_\infty \sqrt{||(\tilde{A}^\dagger \tilde{A} -\tilde{B}^\dagger \tilde{B})||_{\infty}}\leq  \chi||\tilde{B}||_\infty \sqrt{||(\tilde{A}^\dagger \tilde{A} -\tilde{B}^\dagger \tilde{B})||_{F}}
\end{align}
In the second line we have used that, for $X, Y>0$, $||\sqrt{X}-\sqrt{Y}||_{\infty}\leq \sqrt{||X-Y||_{\infty}}$~\cite{bhatia2013matrix}.  Combining\eqref{eq:difference} and \eqref{eq:intermediate}, we arrive at
\begin{align}
||\tau_{AB}^M-\tau_{BB}^M||_1&\leq \chi^2 {\rm max}(1,||\tau_{BB}||_{\infty})||\tilde{B}||_{\infty}\sqrt{\Lambda(q)} e^{-\alpha q/2} \sum_{k=0}^{M-1}||\tau_{AB}^{M-1-k}||_{\infty}=:C(q)e^{-\alpha q/2}\sum_{k=0}^{M-1}||\tau_{AB}^{M-1-k}||_{\infty} \,.
\end{align}
As a last step, we write
 $\tau_{AB}^{M-1-k}=\tau_{BB}^{M-1-k}+(\tau_{AB}^{M-1-k}-\tau_{BB}^{M-1-k})$. Using $\tau_{BB}^{M-1-k}=\tau_{BB}$ for $k\neq M-1$, we get
\begin{align}
\sum_{k=0}^{M-1}||\tau_{AB}^{M-1-k}||_{\infty}&=\sum_{k=0}^{M-1}||\tau_{BB}^{M-1-k}+(\tau_{AB}^{M-1-k}-\tau_{BB}^{M-1-k})||_{\infty} \leq  M\max(1,||\tau_{BB}||_{\infty})\nonumber\\
+&\sum_{k=0}^{M-1}||(\tau_{AB}^{M-1-k}-\tau_{BB}^{M-1-k})||_{\infty} 
\leq  M\max(1,||\tau_{BB}||_{\infty})+\sum_{k=0}^{M-1}||(\tau_{AB}^{M-1-k}-\tau_{BB}^{M-1-k})||_{1}\,,
\end{align}
and so
\begin{align}\label{eq:inequality}
||\tau_{AB}^M-\tau_{BB}^M||_1\leq \tilde{C}(q) M e^{-\alpha q/2}+\tilde{C}(q) e^{-\alpha q/2}\sum_{k=0}^{M-1}||(\tau_{AB}^{k}-\tau_{BB}^{k})||_{1}\,,
\end{align}
where $\tilde{C}(q)=\max(1,||\tau_{BB}||_{\infty}) C(q)$. Let us define
\be
\varepsilon_q=\tilde{C}(q)M e^{-\alpha q/2}\,,\quad \delta_q=\tilde{C}(q)e^{-\alpha q/2}\,.
\ee
By iterating~\eqref{eq:inequality}, we obtain
\begin{align}
||\tau_{AB}^M-\tau_{BB}^M||_1&\leq \varepsilon_q+\delta_q \sum_{k=0}^{M-1}||(\tau_{AB}^{k}-\tau_{BB}^{k})||_{1}\leq \varepsilon_q+\delta_q \{\tilde{C}(q)(M-1)e^{-\alpha q/2}+ (1+\delta_q)\sum_{k=0}^{M-2}||(\tau_{AB}^{k}-\tau_{BB}^{k})||_{1}\}\nonumber\\
&\leq \varepsilon_q+\delta_q \{\varepsilon_q+ (1+\delta_q)\sum_{k=0}^{M-2}||(\tau_{AB}^{k}-\tau_{BB}^{k})||_{1}\}\nonumber\\
&\leq \varepsilon_q+\delta_q \{\varepsilon_q+\varepsilon_q(1+\delta_q)+(1+\delta_q)^2\sum_{k=0}^{M-3}||(\tau_{AB}^{k}-\tau_{BB}^{k})||_{1}\}  \nonumber\\
&\leq \ldots\leq \varepsilon_q+\delta_q \{\varepsilon_q+\varepsilon_q(1+\delta_q)+\varepsilon_q(1+\delta_q)^2+\ldots \varepsilon_q(1+\delta_q)^{M-2}\}\label{eq:almost_done}\\
&\leq \varepsilon_q+\varepsilon_q ^2(1+\delta_q)^{M-2}=\varepsilon_q+\varepsilon_q ^2\left(1+\frac{\varepsilon_q}{M}\right)^{M-2}=\varepsilon_q+\varepsilon_q^{2}e^{\varepsilon_q}(1+O(\varepsilon_q/M))=O(\varepsilon_q)\,,\label{eq:finished}
\end{align}
where, in order to go from~\eqref{eq:almost_done} to \eqref{eq:finished} we have used $\varepsilon_q=M\delta_q$. Putting all together, we obtain
\be
|\braket{\psi_M|\varphi_M}-1|=O(\varepsilon_q)=O(Mq^{(\chi^2-1)/2}e^{-\alpha q/2})\,.
\ee
\end{widetext}
Then, choosing $0<\beta<\alpha/2$, we have that for $q$ sufficiently large $q^{(\chi^2-1)/2}e^{-\alpha q/2}< e^{-\beta q}$,  so finally
\be
|\braket{\psi_M|\varphi_M}-1|=O(Me^{-\beta q})\,,
\ee
which concludes the proof.

\section{LME and stabilizer states}

In this section, we show that any stabilizer state $\ket{\phi}$ gives rise to a deterministic unitary transformation, following the protocol explained in the main text. Given $U\in\mathcal{U}$, let us first recall the definition of Choi-Jamiolkowski (CJ) state~\cite{nielsen2002quantum} as
 \be
 |R\rangle_{s,a} = (U\otimes \Id ) |\Phi\rangle_{s,a}
 \ee
where $|\Phi\rangle_{s,a} = \otimes_{i=1}^M |\Phi^+\rangle_{s_i,a_i}$
and $|\Phi^+\rangle_{s_i,a_i}$ is the maximally entangled Bell state between the spin and the corresponding ancilla at a site $i$. For any $|\psi\rangle\in H$, the action of $U$ can be easily recovered from the knowledge of $\ket{R}$. To see this, let us introduce another ancilla at each site, and prepare this new ancillary system in the state $|\psi\rangle_{a'}$. We have
 \be
 \label{Choiaction}
U|\psi\rangle_s = d^{M} \;_{a,a'}\langle \Phi^+ |R\rangle_{s,a}\otimes |\psi\rangle_{a'}.
\ee
This identity has a very simple interpretation: if we measure both ancillas at each site in a basis containing the state $\ket{\Phi^+}$, and the result of the measure is $\ket{\Phi^+}$ everywhere, then we recover the action of $U$ and the spins are in the state $U|\psi\rangle_s$ after the measurement. Note, however, that this will only occur with a finite probability and thus the action is only accomplished probabilistically in this way.

A LME state~\cite{kraus2009local} naturally gives rise to a Choi state, because a necessary and sufficient condition for a state $\ket{R}$ to be the Choi state of a unitary $U$ is that ${\rm tr}_s(\ket{R}\bra{R})=\openone_a$~\cite{wolf2012quantum}. In order to show that any stabilizer state $\ket{\phi}$ gives rise to a deterministic unitary transformation, we need to make sure that \textbf{(i)} the state $\ket{\phi}_s$ is LME; \textbf{(ii)} one can choose $V\in {\cal LU}$ s.t. $V_{sa}(\ket{\phi}_s\otimes \ket{\bf 0}_a)=\ket{R}_{s,a}$ and $\ket{R}$ is the CJ state of a Clifford unitary transformation. Point (i) was shown in~\cite{kraus2009local}, while (ii) follows from the fact that any stabilizer state is equivalent, up to LU, to a so-called graph state~\cite{schlingemann2001stabilizer,van2004graphical}. Indeed, for a graph state $\ket{\phi}$, it was shown in \cite{kraus2009local} that the unitaries $u_i$ that entangle $\ket{\phi}$ with the ancillas can be chosen to be control-$z$ operators, i.e. $C^z= \openone \otimes |0\rangle\langle 0|+\sigma^z\otimes | 1\rangle\langle 1|$, with the ancilla being initialized in the state $\ket{+}=(\ket{0}+\ket{1})/\sqrt{2}$. In this representation, using the results of~\cite{kraus2009local}, one can explicitly write down the unitary encoded in~\eqref{Choiaction} as
\be\label{eq:clifford}
U = \sum_{\{i_k\}} (\sigma_1^z)^{i_1} \cdots (\sigma_M^z)^{i_M}  \big( \prod_{e } C_{e}^z \big)  \ket{+}^{\otimes M} \bra{i_1 \ldots i_M}
\ee
where $e$ runs over the edges of the corresponding graph associated with the graph state $\ket{\phi}$. It is then easy to see that $U$ is a Clifford operator. Indeed, by a direct calculation we see that it maps $\sigma_k^x \mapsto \sigma_k^z$ and $\sigma_k^z \mapsto \sigma_k^x \prod_i \sigma_i^z $, where the index $i$ runs over all sites that are connected to site $k$ in the graph representation of the state.

\section{The Toric-Code Unitary}

Here we provide more details about the construction for the unitary operator generated by the TC, $U_{\rm TC}$. To this end, we recall that, since $\ket{\rm TC}$ is a stabilizer state, there exists $V_0\in {\cal LU}$ such that $\ket{\Psi}=V_0\ket{\rm TC}$ is a graph state~\cite{van2004graphical}. Hence, it follows from the results of Ref.~\cite{kraus2009local} that
\be\label{eq:r_tc}
\ket{R}=\otimes_j C^z_j (\ket{\Psi}_s\otimes \ket{+}_a^{\otimes M})
\ee
is a maximally entangled state wrt to the bipartition spins-ancillas, where $C^z_j$ is a control-$z$ operator acting on each spin-ancilla pair. Since $\ket{R}$ corresponds to a Clifford unitary operator, cf.~\eqref{eq:clifford}, it can be implemented deterministically, and defines a unitary operator $U_{\rm TC}\in {\cal U}$ whose action is encoded in~\eqref{Choiaction}. From Eq.~\eqref{eq:r_tc}, it is immediate to compute 
\be
U_{\rm TC}\ket{{\bf 0}}_s= \ket{\Psi}= V_0 \ket{\rm TC}\,,
\ee
as anticipated in the main text. This implies in particular that $U_{\rm TC}\notin {\cal QC}_{\ell}$ for $\ell<N/4$.

\end{document}